# Report of the Kavli-IAU Workshop on Global Coordination

# Probing the Universe from far-infrared to millimeter wavelengths: future facilities and their synergies

Workshop held March 26-28, 2024 at Caltech, Pasadena, CA, USA

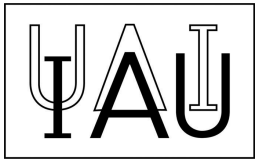
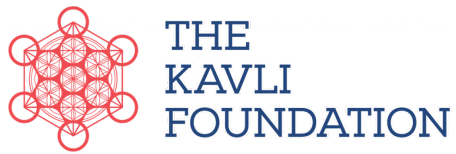

# Table of Contents

# Executive Summary

The Kavli-IAU Workshop on Global Coordination convened 70 researchers from six continents and sixteen countries.  Featuring 14 invited talks and 5 panel discussions, it aimed to define the needs and potential synergies for different facilities at wavelengths from ~30μm to a few cm in the 2030+ era, considering both financial and programmatic limitations, and to explore how to maximize the scientific insights from the data they will yield in the coming decades.  Here we highlight the recommendations that emerged from the workshop; further discussion and elaboration can be found in the main report.

# Major Recommendations

1. ALMA should advance the development, vetting and dissemination of an ALMA2040 vision and roadmap document.

2. Science agencies and implementing organizations should maintain the momentum and schedule towards the construction of ngVLA in this decade. They should continue to work at developing international partnerships to support ngVLA construction and enable wide access.

3. A space-based FIR observatory should be pursued with urgency to avoid losing critical scientific, technical and industrial expertise, and to fully exploit synergies with JWST and ALMA. The international community should remain engaged in the APEX process so their participation results in a more powerful mission.

4. Continued study of large-aperture, wide-field mm/submm telescopes should be pursued to further advance compelling science cases and synergies with other facilities, to advance technology readiness and to develop reliable cost estimates.

# Additional Recommendations

5. Robust programs of technology development for FIR-to-cm instruments should be pursued to preserve technical and scientific expertise and train new experts, achieve the potential of new detectors and signal processing systems and develop new detector technologies.

6. The cost of data systems and data analysis, often considerable, should be built into the project plan and budget of observing facilities from the outset.

7. Students and early-career researchers capable of bridging between data science and astrophysics should be recruited to analyze data, extract science, and create software tools for these purposes. Developing countries should be brought into this process of recruitment and training.

8. Time allocation policies and processes should be evolved to respond to the evolving needs of the science, and enable both coherent large programs as well as innovative, nimble time dependent observing, which will yield more science from the facilities.

9. Science agencies and the community should continue to support the gathering and curation of laboratory astrophysics data relevant to studies in the FIR-to-mm spectral range.

10. Sustainability should be taken into consideration from the outset when designing new observing facilities, with goals and metrics appropriate for each location.

## Scientific Organizing Committee:

Alberto Bolatto, U. Maryland
Ilse Cleeves, U. Virginia
Daniel Dale, U. Wyoming
George Helou, Caltech/IPAC
Kentaro Motohara, NAOJ
Pat Roche, U. Oxford
Linda Tacconi, MPE Garching
Ewine van Dishoeck, U. Leiden
Jonas Zmuidzinas, Caltech

# 1. Preamble on IAU and Global Coordination WG

Progress in astronomy and astrophysics is driven by the continuous improvements in observational facilities and their instrumentation, across the entire electromagnetic spectrum and now also with multi-messenger probes. The largest telescopes and missions almost always involve international partnerships to make them happen. Global cooperation and collaboration are therefore increasingly important since the costs to build forefront telescopes and operate them exceed those that can be afforded by a single nation or region. International strategic planning is essential to explore how such partnerships can be built and how joint projects can be developed, with the goal to maximize the scientific return from these facilities.

To stimulate such collaborations and planning, the International Astronomical Union (IAU) established in 2016 the [Working Group on Global Coordination of Ground and Space Astrophysics](). Its importance is highlighted by the fact that it reports directly to the IAU Executive Committee (EC). A key part of the WG activities is to organize meetings that foster long-term strategic thinking. The WG therefore coordinates one Focus Meeting at every General Assembly covering a broad range of issues, and one focused workshop in between GAs. Previous workshops have centered on [Future space-based ultraviolet-optical-infrared telescopes]() (Leiden, 2017) and on [International co-ordination of multi-messenger transient observations in the 2020s and beyond]() (Cape Town, 2020). The topic for the 2024 workshop was chosen to be the [Far-Infrared and millimeter-centimeter wavelength regime](). It took place March 26-28, 2024 in Pasadena under the excellent organization by George Helou and Jonas Zmuidzinas, and generously supported by the Kavli Foundation.

The goal of this Kavli–IAU workshop was to examine the needs and requirements of different observing facilities in the far-infrared to millimeter and centimeter wavelengths from 2030 onwards (taking financial and programmatic constraints into account). The workshop also considered the synergies and complementarities among these facilities, and explored how to maximize the scientific insights from the data they will yield. The workshop was timely in light of the US Astro2020 recommendations, the new NASA Astrophysics Probe Explorer (APEX) program as well as developments in Europe and Asia on strategic planning for ground and space-based astrophysics. Note that the IAU facilitates these discussions and aims to stimulate long term plans, but it does not endorse any particular mission or facility more or less than another.

# 2. Introduction

## 2.1. Astronomy at far-infrared and mm-cm wavelengths

Astronomy is conducted across the whole electromagnetic spectrum, with different classes of objects and physical processes dominating in different wavelength bands. The far-infrared and mm wavelength regions are where the peak output of cool objects occurs, including planets and star-forming clouds as well as the dust and gas that enshrouds some stars and galaxies. Numerous unique atomic, molecular and solid-state diagnostic spectral features lie in this wavelength region and can be used to probe both warm and cold gas to significantly advance our understanding of galaxy-, star- and planet formation. For the purposes of this workshop, the wavelength regime covered is roughly 25 μm (where JWST ends) to 1 cm (the shorter wavelength part of the proposed ngVLA), or in frequency space 20 - 9000 GHz.

***Far-Infrared (FIR):*** The Earth's atmosphere imposes severe restrictions on ground-based FIR measurements, as the atmosphere is opaque between 25 and 300 μm. Furthermore, telescopes on the surface of the Earth emit infrared light characteristic of their temperatures, which adds to the "background" emission from the atmosphere to swamp the faint astronomical signals. Stratospheric platforms can be used but atmospheric absorption and emission remain significant. These factors require sensitive FIR facilities to be placed above the atmosphere and cooled to low temperatures to mitigate this background. In turn, this places severe constraints on the telescope apertures of feasible FIR facilities. A number of satellite missions between 1983 and 2013 opened up observations in the FIR and provided rich data sets that have revealed many new phenomena. The early missions used cryogenically-cooled telescopes (2-6 K) with modest apertures (0.6 – 0.9m) followed by the 3.5m, 80 K Herschel Space Observatory which operated between 2009 to 2013. No FIR space missions have launched since 2013 and none are under development today, extending this hiatus to at least two decades. New facilities operating at FIR wavelengths are urgently needed to build further on the legacy of the earlier missions, matching the scientific capabilities at other wavelengths and exploiting the advances in detector technology that have the potential to achieve orders of magnitude improvements in the sensitivity of FIR measurements.

Instruments attached to telescopes on high-altitude balloons or flown in aircraft have made significant contributions to the development of FIR instruments and have an important role in technology development and testing. But the only general use FIR observatory (SOFIA) is no longer in operation, and balloon-borne platforms cannot provide the sustained, sensitive observations needed to realize the potential of the performance of current detectors.

***Millimeter/Submillimeter (mm/submm)*:** Transmission windows in the Earth's atmosphere begin to open up at 5,000 m altitude sites at wavelengths beyond 300 µm (0.3 mm, frequencies below 1 THz). Deep absorption bands, primarily due to water vapor, mean that high and cold sites are necessary for mm observations, especially at higher frequencies, providing the low pressure and dry conditions required for maximum transmission at wavelengths between 0.3-1 mm. Single aperture telescopes and interferometers operating in the mm and submm are therefore located on high mountains, with the bulk of the spectroscopic observations done at 0.8-1 mm. The high spatial resolution provided by interferometers coupled with the high spectral resolution allowed by their heterodyne receivers have produced transformational scientific discoveries, complementing the wider-field capabilities of the single dish facilities. The international ALMA observatory is the pre-eminent mm and submm facility today, with receivers spanning wavelengths of 0.3-7 mm.

Many of the world's leading mm-wave telescopes can be linked together across continents to enable very long-baseline interferometry, providing even higher spatial resolution and enabling exciting other science investigations such as imaging of black holes, first achieved with the Event Horizon Telescope (EHT).

***Centimeter (cm)*:** While the atmosphere still provides limits to observing at wavelengths of a few mm, at centimeter wavelengths the atmospheric transmission approaches 100% from most observing sites, allowing both single-dish and interferometer observatories either to offer general-purpose capabilities or optimize their design to a specific science goal such as the Cosmic Microwave Background but extract much ancillary science from the same data. The JVLA and its proposed ngVLA follow-on lead the field of general-purpose interferometer observatories in the 1-116 GHz range. The longer wavelengths being planned by the Square Kilometer Array (SKA) were outside the scope of this Workshop.

## 2.2. Science Drivers

The physical conditions and structure of interstellar and circumstellar media (ISM and CSM) dictate where and how stars and planets can form. The feedback effects from both star formation and black hole driven activity are important processes in galaxy evolution that have a dramatic effect on the physics of the ISM. Important cooling lines of the ISM and key atomic, molecular and solid-state diagnostics are found in the FIR to mm wavelength region of the spectrum. This is also where the thermal emission from cold dust, and more generally the energy output from star-forming clouds and galaxies, peaks. These wavelengths therefore provide unique probes of relatively cool, dense interstellar material, central to the study of forming stars, protoplanetary disks, and young forming exoplanets. On larger scales they probe dust and dense neutral gas in the ISM of galaxies and around highly obscured accreting supermassive black holes. This spectral region thus offers key diagnostics of the physical conditions that are important for further progress across a broad canvas of astrophysics, and benefits from drastically reduced dust extinction relative to optical/near-IR observations, allowing us to probe the densest ISM regions.

The FIR to mm region is also particularly rich in lines of a wide variety of molecules in gaseous and solid form, from simple species like water to increasingly complex molecules such as sugars, ethers, cyanates, and aromatic hydrocarbons. Following the trail of molecules from clouds to planet-forming disks and mature planets will elucidate the role they play in the emergence of life elsewhere in the universe. This is also the spectral range into which mid-IR emission shifts in studies of the early universe. A more detailed description of the science drivers can be found in Section 5.

## 2.3. Goal and Timing of this Workshop

Planning is underway for a new generation of facilities operating from the FIR to cm wavelengths. The James Webb Space Telescope is now operating, probing the mid-IR out to 30 μm, hopefully for the next 10-20 years. ALMA continues to operate at submm to mm wavelengths, and will be upgraded by 2030, but there are no plans yet for further enhancements in the 2030+ era. Following on from the pioneering work of the Infrared Astronomical Satellite (IRAS), the Infrared Space Observatory (ISO), the Spitzer Space Telescope, Akari and the Herschel Space Observatory, there is great interest across multiple space agencies in a probe class/medium class FIR space mission, and multiple mission concepts are being developed, but none has moved beyond the concept phase. The ngVLA is a proposed mm-cm class facility that aims for a tenfold improvement over the sensitivity of the JVLA. Although the ngVLA will overlap with ALMA's lowest frequency bands, the current plans call for nearly 10x larger collecting area along with substantially longer baselines (~30x), yielding milli-arcsecond angular resolution.

This planning is benefiting from major technological advances in the last decade which enable or enhance several detection methods at different spectral ranges, which improve substantially the cost of facility construction, or which enable novel architectures for telescopes on the ground and in space. It is also informed by the outcome of major strategic planning activities by the US National Academy of Sciences (Astro2020, or the Decadal Survey of Astronomy and Astrophysics), by the European Space Agency (Voyage 2050, or ESA's Science Programme up to 2050), and by equivalent organizations in several countries supporting major observing facilities in this spectral range.

This Workshop was convened to examine the needs and requirements for different facilities at wavelengths from FIR to cm in the 2030+ era taking financial and programmatic constraints into account, to consider the synergies and complementarities among these facilities, and to explore how to maximize the scientific insights from the data they will yield. The attendance comprised 70 researchers from six continents, and included scientists, observatory leaders and science agency representatives.  It featured plenary sessions with 14 invited talks and 5 panel discussions as well as break-out sessions for more focused topics. The Science Organizing Committee met on the last afternoon to summarize the main findings and recommendations developed by the attendees for this Report.

# 3. Facilities Landscape: Findings and Recommendations

## 3.1. ALMA

The Atacama Large Millimeter/submillimeter Array, ALMA, is a collection of 66 antennas operating as an imaging interferometer at wavelengths ranging from 7 mm to 350 µm (roughly 40 to 860 GHz). ALMA is a partnership between North America, Europe and East Asia, in cooperation with Chile that has been delivering high quality observations for a decade, and has demonstrated the benefits of global collaboration in producing advances in capability that are transformational and beyond the resources of individual countries or even continents. The contributions of technology, skills and funding from all partners have been and continue to be essential in delivering the full capability of the facility.  The ALMA site at 5000m elevation on the Chajnantor plateau in northern Chile provides exceptional conditions for sub-millimeter astronomy, and opens up the highest frequency bands that cannot be accessed from lower elevations.  The excellent conditions enable routine observations at the scientific "sweet spot" work-horse wavelengths of 0.8-1 mm. The longest baselines currently available (16 km) provide high spatial resolution (18 milli-arcsec at 1 mm)  while the large number of antennas enables high fidelity images. Full exploitation of the highest frequency capabilities at 0.3 mm requires sufficient agility to adapt to variable weather conditions and may require adjusting scheduling priorities.

Even higher spatial resolution observations of less than 20 micro-arcsec at 1 mm are enabled by very long baseline interferometry (VLBI) combining ALMA with other mm telescopes around the world, opening up unique high profile science studies. ALMA's large collecting area and sensitivity are key in providing high quality mm-VLBI images.

The demand for observations with ALMA has been very high since the first observing cycle, with oversubscription factors of 10 or more for many configurations. Astronomical communities around the world have exploited ALMA's images and spectra and the data archive, with the demand projected to increase further in coming years.  A program to increase ALMA's sensitivity and bandwidth, the wideband sensitivity upgrade (WSU), is underway and expected to conclude around 2030. The WSU has been approved by the ALMA Board using the development funding included within the contributions under the ALMA agreement, which has provided the framework and stability to develop this decade-long programme.

The ALMA community has started to discuss a vision for ALMA in the 2040s with multiple scenarios of enhanced capabilities under consideration.  These scenarios broadly address improved sensitivity, longer baselines for better angular resolution, and increased fields of view, along with improvements to the data system and archive.

This Workshop was charged with considering future upgrades and enhancements to the facilities beyond WSU.  A number of potential enhancements have been proposed including increasing its collecting area, extension to longer baselines and the development of multi-pixel receivers.  Discussions during the Workshop, including the break-out discussion dedicated to high resolution science, led to these **findings**:

1. **ALMA has been a hugely successful international collaboration, implementing an observatory with transformational capabilities, and yielding remarkable scientific results across all of astrophysics and Solar System studies.**
2. **Smaller, but still powerful facilities, such as NOEMA and the SMA have played and will continue to play a critical role in technology development for future ALMA upgrades.**
3. **Compelling science cases exist (e.g. planet formation and distant galaxies; cf. Section 5) that would require an order-of-magnitude increase in the spectral line sensitivity of ALMA, which would necessitate an expansion in the collecting area and a commensurate boost in construction, operations and maintenance funding.**
4. **Increasing ALMA's sensitivity is the first priority for almost all science by the community, as sensitivity is key to observing fainter transitions and for expanding the utility of higher angular resolution data. The planned WSU** will increase the ALMA line sensitivity at most modestly, so that increase in sensitivity will remain the first priority.
   5. **Particularly strong scientific cases exist for upgrading the sensitivity in ALMA bands 6 and 7 (or between 0.8 and 1.4 mm).**
   6. **It may be possible that use of an antenna design that is simplified relative to the current ALMA antennas — relaxing surface accuracy, slew speeds, and other requirements as well as populating them with fewer receiver bands — can lower the cost of populating existing antenna pads and reduce the cost of achieving a 3x to 10x improvement in line sensitivity.**
   7. **Separately, a baseline expansion within the current astronomy concession (yielding a factor of ~2 longer baselines) should still be pursued and is probably doable within the existing budgetary envelope of the ALMA upgrade plan.**

Based on these findings, we **recommend** that the ALMA partners (and their communities):
   1. **Advance the development, vetting and dissemination of an ALMA2040 vision and roadmap document.**
   2. **Develop a comprehensive community engagement plan, for instance organizing international workshops and setting up working groups to develop white papers that further explore the scientific and technical aspects of (up to) an order-of-magnitude increase in line sensitivity, or other potential facility enhancements.**
   3. **Investigate the possibility of advancing long lead time items such as a simplified antenna design capable of performing at frequencies up to band 7 at the ALMA site (for example, building on the lessons learned from the ngVLA antenna development).**
   4. **Continue to fund, operate and upgrade mid-size interferometers in the Northern Hemisphere.**
   5. **Consider the milestones in the funding landscape over the rest of this decade — when ESO will be looking for the "Next Big Facility", the US will be having a new Decadal survey, and Japan developing new astronomical priorities — as well the possibility of bringing in new partners to fund such a major upgrade.**

## 3.2. ngVLA

The next-generation Very Large Array, ngVLA, is one of the top three priorities for ground based astronomy in the US 2020 Decadal Survey. It will operate at wavelengths ranging from 21 cm to 2.6 mm, replacing and building on the remarkable successes of the aging Jansky Very Large Array (JVLA) and the Very Large Baseline Array (VLBA). It is planned to comprise 244 antennas with 18 m diameter distributed around a core concentration on the Plains of San Agustin JVLA site in New Mexico, and baselines extending across the continental US, northern Mexico, and out to Hawaii and Puerto Rico. Its powerful capabilities respond to ambitious science drivers, from imaging terrestrial planet formation and determining the astrochemical initial conditions for planetary systems and life, to testing gravity and studying the formation and evolution of supermassive black holes and galaxies. The project, led by the U.S. National Radio Astronomy Observatory (NRAO), is currently a U.S. National Science Foundation (NSF) Major Research Equipment and Facilities Construction (MREFC) Design Candidate. As such, it will undergo a series of three design reviews before formal consideration for construction. The project office is currently funded through the completion of the first two of these full-system design reviews, with the NSF-run Conceptual Design Review (CDR) being held later this year (September 2024) and its Preliminary Design Review (PDR) to be held in early 2026. Upon the successful completion of these reviews, additional funding will be requested to complete the final design by mid-2028 so that construction can commence in 2029 if funding is provided in time, which would satisfy the Astro2020 Decadal Survey recommendation for project construction to begin this decade. During this time, the ngVLA project will continue to identify and formalize both U.S. inter-agency and international partnerships. In parallel, the ngVLA antenna design and prototyping effort is being funded independently through the NSF Mid-Scale Innovation Program (MSIP). The antenna design effort was completed in late 2022 by the contractor *mtex antenna technology GmbH*, and the prototype is now being constructed in Germany and Spain. Antenna components will begin to arrive at the JVLA site in June 2024, where a previous ALMA antenna test pad has been retrofitted to accommodate the ngVLA prototype design, with full assembly to be completed in October 2024 and the handing over to NRAO for scientific validation to occur in February 2025.

Although as a Decadal Survey priority there is every expectation that the ngVLA will be constructed, maintaining the interest and vigorous support of the user community while that happens is extremely important. The Workshop developed the following **findings**:
1. **The science case for the ngVLA is highly compelling, and encompasses key science that cannot be carried out with any other existing or planned facility, for example, the formation of terrestrial exoplanets.**
2. **Without an ngVLA there would be a large gap in sensitivity and resolution at frequencies of 20 GHz to 116 GHz. ALMA, despite its coverage starting at 35 GHz, has 10 times smaller collecting area and 20-200 times shorter baselines.**
3. **NSF and NRAO have placed ngVLA on a path to construction following the U.S. Decadal Survey recommendation.**

4. **The ngVLA drives technology development that has very significant synergies with a possible ALMA upgrade as described above, in particular for antennas, receivers, and correlators.**

Based on these findings, the Workshop offers the science agencies and implementing organizations responsible for the ngVLA these **recommendations**:
1. **Maintain the momentum and schedule towards the construction of the ngVLA in this decade.**
2. **Maintain and expand the engagement of the relevant international user community, which is heavily shared with ALMA, with an eye toward the sustained development and updating of the ngVLA science case in order to support the ngVLA design and plan its operation.**
3. **Continue to work at developing strong international partnerships to support the ngVLA construction (and to provide margin against budget constraints), and to enable wide access to the ngVLA with the goal of producing the best possible science.**

## 3.3. Space Far-Infrared

The FIR provides the crucial coverage between the mid-infrared science enabled by JWST or mid-IR survey facilities and longer wavelength science carried out with ALMA and the JVLA and eventually ngVLA. Despite the exciting science that the FIR uniquely enables, the community had no access to these wavelengths from a space based platform for more than ten years. Since Herschel was decommissioned in 2013, sub-orbital platforms such as high altitude balloons and the now-retired SOFIA airborne observatory have been the primary access points to the FIR sky. Neither, however, can compete with a dedicated space platform, which provides high sensitivity, broad wavelength coverage, and the mission duration to carry out large campaigns to observe many targets. To address this gap, the community developed two FIR space mission concepts, SPICA led by Europe and Japan ([Roelfsema et al. 2018](#)) and the Origins Space Telescope (OST) in the US during preparations for the US Astro2020 decadal ([Meixner et al 2019](#)). The technological leap forward required for these missions was a large, cryogenic mirror paired with detectors sufficiently sensitive to exploit the greatly reduced background. SPICA was canceled in the middle of the development process, and while OST was recommended by Astro2020, it was not identified as the first NASA Flagship class mission to go forward.

The most promising opportunity on the horizon, however, also came out of the US Astro2020 decadal survey in recognition of the space FIR gap after the cancellation of SPICA. Astro2020 recommended that the community develop a "Probe-class" mission in the FIR or X-ray wavelength regime. This recommendation led to NASA announcing an opportunity for an "Astrophysics Probe Explorer" with a $1B cost-cap excluding launch. Of this cost, up to 30% can be raised external to NASA. The expected launch timing is the early 2030s, requiring all of the components to already have high technological readiness, which has influenced the design of the on-going concept planning. We commend NASA's rapid response to the community's call for this crucial gap in scientific capabilities for the field at large.

The successes of Spitzer, Herschel, and now JWST have generated great enthusiasm among early-career scientists for space FIR astrophysics. This enthusiasm is heightened by the conspicuous FIR gap wedged between two very powerful facilities, ALMA in the submm and JWST in the mid-IR.

Given the breadth of science that the FIR enables, three FIR specific mission concepts were proposed to the APEX opportunity and are under consideration as of May 2024. Each of these is tuned to optimally carry out different aspects of FIR science. Without a larger NASA Flagship FIR mission, it is clear that such tradeoffs are necessary to meet the cost cap. A clear and detailed overview of the pros and cons of each of the four mission concepts being considered as of two years ago is provided in [van der Tak et al](). (2022), and thus we only summarize the salient points here and direct the reader to this overview for additional background context. The three current APEX missions under consideration, alphabetically listed, are FIRSST, PRIMA, and SALTUS.

**FIRSST**, the Far-Infrared Spectroscopic Survey Telescope, is designed around a cryogenic 2m-class telescope. While smaller than Herschel's 3.5m mirror, new detector technology combined with an actively cooled mirror allow this probe to be several orders of magnitude faster on broadband and spectral science. FIRSST features a low resolution wide-bandwidth spectrograph, a heterodyne spectrograph, and virtual phased arrays (VIPAs) modules. VIPAs provide high resolution (R~100,000) over a predetermined spectral range, and enable sensitive spectroscopy centered on lines of interest such as ground-state HD, [CII], and [OI].

**PRIMA**, the PRobe Infrared Mission for Astrophysics, is also a cryogenic 2m-class telescope offering similarly large improvements in sensitivity over Herschel. PRIMA features a hyperspectral imager (R~10) with wide wavelength coverage and polarimetric capabilities at long wavelengths. PRIMA also has a long-slit low resolution (R~100) spectrometer that has the sensitivity to rapidly map total line fluxes over a large bandwidth (24-235 $\mu$m), along with a medium resolution mode with tunable resolution (R~4,400 x 112 $\mu$m/λ) provided by a Fourier transform spectrograph (FTS) module. This flexible payload enables wide-bandwidth science, ranging from planet and star formation studies to galaxy evolution and the interstellar medium.

**SALTUS**, the Single Aperture Large Telescope for Universe Studies mission concept, relies on an inflatable 14 m dish, seven times the size of the previous concepts. While it cannot be cryogenically cooled, the size of the dish gives it a similar sensitivity to PRIMA and FIRSST. A key feature of SALTUS is its high angular resolution. SALTUS also has both low resolution (R ~ 300) and heterodyne high resolution (R ~ $10^6$ - $10^7$) modes for studying both solids (dust and ice) as well as gas in disks, with the possibility of imaging gas and ice tracers within individual disks.

All three mission concepts are enabled by recent advances in superconducting detector technology and use of arrays of kinetic inductance detectors (KIDs) for their direct-detection (i.e., non-heterodyne) instruments. Investment in developing this technology over the past two decades has led to construction of large ground-based instruments (e.g., [NIKA 2](#) on the IRAM 30 m telescope), and KIDs have now reached the sensitivity levels needed for FIR space missions with cryogenic telescopes. For example, [Day et al.](#) (2024) recently reported KIDs capable of counting single photons at $\lambda$=25 $\mu$m.

The Workshop discussions demonstrated that the community is excited for all of these missions, and will make the best use of whatever facility is selected for development. Given the uncertain funding landscape, and the internal competition within APEX against X-ray missions, the community recognizes the importance of remaining positive and supportive, even as these missions are weighed against one and others during the lengthy APEX review process. The collegial spirit of the community clearly extended to enthusiastic early career scientists, brought in through their engagement with complementary facilities like ALMA and JWST, who expressed strong support for sensitive FIR missions.

Workshop discussions resulted in the following **findings:**
1. **There are unique, highly compelling science cases stretching across many fields from galaxies at cosmic noon to solar system objects, and very strong community support for FIR wavelength science, especially in combination with ALMA and JWST.**
2. **The FIR science promise was recognized by the US-based Astro2020 Decadal Survey, which recommended a FIR or X-ray Probe-class mission. NASA quickly implemented the Probe recommendation as the APEX program, which motivated the community to submit three exciting FIR mission concepts.**
3. **There has been no space FIR observatory since 2013, and this absence could extend by two more decades. This translates into a mounting risk of** **completely losing critical technical and scientific expertise, infrastructure and industrial capacity in this unique part of the spectrum.**
4. **Technological progress enables powerful new missions even when budget considerations dictate difficult design choices in other major aspects.**
5. **Bridging this wavelength gap is urgent given the limited lifetime of synergistic facilities at shorter wavelengths, like JWST.**

Based on the findings above, we offer the following **recommendations:**
1. **A space-based FIR observatory should be pursued with urgency to avoid losing critical scientific, technical, and industrial expertise as well as to fully exploit the synergies with JWST.**
2. **The strong community endorsement of the need for a space FIR mission should be considered in the first round of NASA APEX selections.**
3. **The international community should remain engaged in the APEX process so it can augment NASA funding for a more powerful mission.**
4. **The whole community should rally around any FIR Probe mission ultimately selected by NASA to help optimize it and eventually use it for their science.**
5. **Space agencies should continue the support of technology development and suborbital missions to improve technological readiness for major FIR future missions, including a future FIR flagship or an L-class mission.**

## 3.4. Single-Dish and Other Facilities

Transformed by ALMA, astronomy at millimeter/submillimeter wavelengths now plays a major role in astrophysics. ALMA's success was built on a foundation of science and technology developed in earlier decades using pioneering smaller facilities, including both interferometers and single-dish telescopes. The science performed using these facilities motivated the construction of ALMA and [the IRAM upgrade to NOEMA](); examples include the fascinating images of the [HL Tau]() protoplanetary disk and the discovery and detailed studies of dusty star-forming ["submillimeter galaxies"]() at high redshift. Together with Herschel-HIFI, these facilities played an essential role in advancing the superconducting ([SIS tunnel junction]()) receiver technology that enables ALMA. In addition, the single-dish telescopes spurred the invention, development, and demonstration of newer technologies such as bolometer array cameras, superconducting detectors, and direct-detection spectrometers. These technological developments fed directly into the SPIRE instrument for the [Herschel Space Observatory](), and alumni of these earlier ground-based projects are now playing leading roles in the Far-IR Probe missions proposed to NASA.

ALMA is an excellent instrument for studies that require high sensitivity, high angular resolution, and/or high spectral resolution. However, with its narrow field of view limited to the primary beam of a 12 m telescope, ALMA is not designed for surveying large areas of sky. Furthermore, heterodyne interferometers are insensitive to flux that is spatially extended over scales comparable to or larger than the primary beam. A brief summary of the features of single dish telescopes and interferometers, together with the current state of detector systems, is provided in Appendix A.

Although interferometers were used in earlier experiments, e.g., [CBI](#) and [DASI](#), cosmic microwave background (CMB) projects now use single-dish telescopes explicitly designed for wide-field imaging and equipped with cameras or polarimeters using multi-kilopixel detector arrays. Examples include the 6 m Atacama Cosmology Telescope ([ACT](#)) and the 10 m South Pole Telescope ([SPT](#)), which are located at excellent sites and are carefully designed to minimize backgrounds. The resulting high mapping speed is necessary for CMB science but also has had a significant impact on mm/submm astronomy. For example, SPT and ACT surveys have identified a [sample](#) of dusty high-redshift galaxies that are gravitationally lensed and are therefore bright, providing important targets for [further study](#) with ALMA or JWST. SPT and ACT surveys have also detected millimeter-wave [transients](#), mostly flaring stars, giving an initial taste of time-domain astronomy at these wavelengths. In the future, larger CMB experiments such as the [Simons Observatory](#) will probe deeper and wider, significantly increasing the discovery space.

Yet, CMB experiments are not optimized for mm/submm astrophysics. They must cover very large fields which dictates use of relatively small telescope apertures with only modest sensitivity for point sources and modest angular resolution, and they operate single-purpose observing programs optimized for CMB science rather than multiple surveys targeting distinct astrophysics goals. They are equipped and optimized for broad-band continuum observations. Such considerations, along with the rapid evolution of focal plane array technology (e.g., using KIDs; Appendix A) argue for a large single-dish mm/submm telescope..

The existing large single-dish telescopes would certainly benefit from stronger support for aggressive development of [array instrumentation](#) and improvements to their [surface accuracy](#). However, one of the fundamental limitations of existing single-dish telescopes is that they were not designed for a wide field of view and most were not designed for submillimeter wavelengths. The antenna surface accuracy limits the performance at shorter wavelengths; for example, the aperture efficiency of the 100 m GBT falls below 30% at 100 GHz. Thus, there have been a number of proposals to construct large, accurate, wide-field single-dish telescopes, including the original [25 m CCAT](#) concept endorsed in the U.S. Astro2010 Decadal Survey, the 50 m Large Submillimeter Telescope (LST) concept developed in Japan, the CMB-HD concept comprising two 30 m telescopes equipped with 2.4 x $10^6$ detectors proposed to the Astro2020 decadal survey, and most recently the EU-funded AtLAST study for a 50 m submillimeter telescope in Atacama. Compared to ALMA/WSU, AtLAST would offer comparable point-source sensitivity but with >1000x faster mapping speed (see Appendix B)[1].

The science goals for AtLAST are documented in a series of recent white papers and span a wide range from solar system science to cosmology and the high-redshift universe. Highlights include deep multiband and spectroscopic galaxy surveys and the potential to study the influence of feedback on the circumgalactic medium. Many of these science goals require both fast mapping and spectral resolving power, which requires spectroscopic focal plane arrays. Such arrays could either use incoherent detection, which provides modest spectral resolution but potentially over very large fields, or coherent detection, which is intrinsically capable of very high spectral resolution. For the former, arrays of superconducting on-chip KID-based spectrometers as illustrated by the DESHIMA or SuperSpec projects offer an attractive solution but will require considerable further investment to fully realize their potential. Ultimately, arrays of spectrometers are envisioned for line intensity mapping experiments, an emerging technique that promises rich returns in astrophysics and cosmology. There is also a growing interest in advancing coherent focal-plane array receiver technology to meet the needs of projects such as AtLAST, pushing formats to $10^3$ feeds or more, but again this will require significant investment. Demonstration platforms are critical for advancing these technologies, so sustaining the operation and development of current submm facilities is an important component of building the case for a future large aperture facility.

Recent decisions to pause development of new astronomical facilities at the South Pole because of infrastructure limitations pose a challenge for future developments of CMB and submm facilities.  Furthermore, after more than 20 years of operation, many of the highest-frequency submm observatories have either closed or face a change in management. Ways need to be found to maintain access to the highest frequency facilities to ensure science programmes, technology developments and test beds for future projects as well as the extension of VLBI, including the EHT, to higher frequencies.

Although the AtLAST team has significantly advanced the case for a large submillimeter telescope, more remains to be done to justify any potential large investment, both in terms of science and technology and the key operating frequencies. In the near term, projects such as the 6 m CCAT-prime will demonstrate substantial advances in KID technology while instruments such as Toltec on the 50 m LMT will advance the AtLAST science case by performing deep, wide surveys in the 1-2 mm range. Opportunities for synergies with other observatories, e.g., a possible Far-IR Probe, should be explored thoroughly. The instrumentation concepts for a large submm single dish telescope also need development and maturation and close integration with the science cases. Yet, construction need not wait for the instrumentation technology to be fully developed – all major observatories have benefited from instrumentation upgrades over their lifetimes.

---

[1] After this Workshop was held, additional funding was obtained to continue development studies of a large single dish with an international collaboration, merging AtLAST with LST, and maturing technology.

**Findings**

**1. Existing small/mid-size single-dish telescopes have been very productive, and their capabilities have increased dramatically over time due to advances in instrumentation (e.g., large array cameras) so that they continue to contribute significantly to mm/submm astrophysics.**

**2. Single-dish telescopes played a key role in developing and demonstrating technologies used for Herschel, ALMA, and the proposed far-IR probes. Their diversity provides unique capabilities and increased agility, and they offer opportunities for new science (e.g., line intensity mapping or time-domain surveys), instrument development, technology demonstration, and training the next generation.**

**3. Existing single dishes are essential for mm/submm VLBI including the [Event Horizon Telescope](), which is also benefiting from the addition of new stations such as the [Africa Millimetre Telescope]() and the [Greenland Telescope]().**

**4. Wide-field surveys from new CMB experiments such as [Simons Observatory]() will also advance mm/submm astrophysics and provide a deeper look at the time-domain sky at millimeter wavelengths.**

**5. Significant capability advances are possible at relatively modest cost through development of advanced single-dish instrumentation on existing telescopes. Spectroscopic array instruments, both direct-detection and coherent, could provide revolutionary advances but will require significant technology development to mature. This will require investment, but the cost is modest compared to a new large observatory.**

**6.  A large aperture (>40m), wide-field mm/submm telescope would provide a fast continuum and spectroscopic mapping capability that is complementary to ALMA's strengths. The community is rapidly developing and demonstrating the necessary instrumentation technology, and the science case for such a facility has also advanced in recent years.**

**Recommendations**

**1.  We encourage further study of large-aperture, wide-field mm/submm telescopes to pursue compelling science cases, to define specifications for the telescope and instrumentation suite, and to make a reliable cost estimate. Funding and governance models should be explored, and technology roadmaps should be developed to aid planning.**

**2.  A vigorous instrumentation technology development effort should be pursued to achieve the potential of new detectors and signal processing systems. Existing single-dish telescopes can be used to demonstrate the advanced instrumentation that results, e.g., spectroscopic arrays, increasing the capability of these facilities and advancing science.**

**3.  Access to high-frequency submm telescopes should be supported to provide technology testbeds, to pursue science programs, and potentially for higher-frequency VLBI.**

# 4. Additional Findings and recommendations

## 4.1. Preservation of Expertise

**Finding:** The technical demands of developing facilities in the FIR-to-cm spectral range can be met only by the unique skills of engineers and technicians specialized in scientific instrument design and construction, spread between the high-tech industry and university or observatory laboratories.  Such expertise cannot be maintained without continual exercise of skills through the building of facilities or the development of technology. With the potential long gaps between facility inceptions, the risk of losing critical expertise can only be mitigated by stable, significant programs of technology development in this spectral range, and continuing opportunities to deploy the technology on existing facilities.

**Recommendation:** Science agencies and the community should partner on robust programs of technology development in the FIR-to-cm spectral range to preserve technical and scientific expertise, and to train new generations of experts. A minimum viable network of facilities should be maintained to allow flexible, agile demonstration of technologies.

## 4.2. Observatory Data Systems

**Finding:** As instruments integrate new detection technologies and increase their spectral resolution and pixel count by leaps and bounds, their output is steadily migrating into the realm of big data. Processing, validating and analyzing these mounds of novel data, then archiving and serving them to the research community are essential activities of the science undertaking. While increasingly daunting tasks with correspondingly high price tags, these activities must be carried out if we are to extract science from the hardware investment. The data system cost can range from under ten percent to a quarter or more of the total project cost, apart from the labor cost of data analysis and science publication.

**Recommendation:** The cost of data systems and data analysis for observing facilities should be considered by the community, the designers and builders and the funding agencies from the outset as an integral part of the project plan and budget, rather than as an afterthought.

## 4.3. Demographic Challenge

**Finding:** The analysis of those mounds of data will fall primarily to students and early career scientists eager to extract science from new facilities, in an extension of a long tradition in astrophysics. They will need new training and new generations of big data tools, some of which can be repurposed from the commercial big data realm, and some will need to be created in response to the novelty of instruments, detectors and data collection. Such early-career, talented researchers needed to meet the challenges will not be readily found using normal approaches to recruiting in astrophysics.

**Recommendation:** Recruiting student and early-career researchers capable of bridging between data science and astrophysics should be pursued with creative approaches beyond current practices in the astrophysics community. These early-career researchers should be trained in analyzing large amounts of data as well as creating software tools addressing the novelty and complexity of instruments and facilities.

## 4.4. Involving Developing Countries

**Finding:** In addressing data analysis challenges, the human resources of developing countries offer a well proven reservoir of computing talent and science enthusiasm. With the internet connecting countries nearly seamlessly, it is relatively easy to organize the participation of that talent in the data analysis challenges.  This participation will also enhance general scientific activities in the developing countries with their established record of accelerating economic and educational development.

**Recommendation:**  Communities supporting large observing facilities should establish collaborations with colleagues and institutions in developing countries aimed at engaging local talent in the data analysis challenges and subsequent scientific exploitation of large facilities.

## 4.5. Observatory Time Allocation Process

**Finding:** Observatories with powerful instruments and many capabilities are already attracting proposals for scientifically rich programs whose data acquisition is more challenging to schedule or to complete in a timely fashion.  Examples are large surveys aimed at homogeneous data collection for unbiased analysis, time-domain programs requiring carefully tailored cadences or rapid followup, coordinated data collection across multiple facilities, or technically demanding observations such as high frequency or polarization with ALMA.  The science and community expectations have evolved rapidly while the Time Allocation Process has progressed more slowly.

**Recommendation:**  Evolve the Time Allocation policies and processes to respond to the evolving needs of the science and community, and to enable both coherent large programs as well as innovative, nimble time dependent observing, which will yield more science from the facilities.

## 4.6. Laboratory Astrophysics

**Finding:** As the observational capabilities improve, the need for more sensitive and accurate lab astrophysics data will escalate in tandem.  As fainter signals at higher spectral resolution over broader frequency bands are collected, it becomes more challenging to identify the transitions in these data without the relevant lab astro measurements on gas phase species, ices and dust in all possible states of ionization, excitation, density or temperature. Once gathered in the lab, these data need to be curated in well organized, consistent and easily explored repositories for efficient access by the community.

**Recommendation:** Science agencies and the community should continue to support the gathering and curation of laboratory astrophysics data relevant to studies in the FIR-to-mm spectral range.

## 4.7. Sustainability

**Finding:** As climate change and mitigation efforts penetrate all segments of society it is appropriate for the astronomy community to participate in the mitigation efforts given the scientific foundation of the issue. Designing mm/submm observatories with sustainability of the power supply in mind is challenging because of the remoteness of their location, but it is a significant factor in engaging both the public and new generations of scientists and engineers to join the observatory development teams.

**Recommendation:** Science agencies and the community should take sustainability into consideration from the outset when designing new observing facilities, with goals and metrics appropriate for the specific circumstances of each location.

# 5. Science landscape, facilities requirements: Findings

The Workshop discussions on facilities were informed by the current state and promise of those facilities, by technology developments and potential, and of course by the major unanswered Astrophysics questions of today and their anticipated extensions, and the data needed to address these questions. These data needs in turn suggest the enhancements and additions to current facilities. This section summarizes the main outstanding questions in representative areas of research and the capabilities required to address these questions in the 2030s and beyond.

**Astrochemistry**:
Molecules are found throughout the Universe, from the nearest planets to the highest redshift galaxies. What is the chemical composition of our Universe? How does the chemical inventory evolve alongside *and help shape* the process of star and planet formation? Is the chemical make-up of interstellar clouds imprinted upon the raw organic inventory delivered to planets or is the chemistry catastrophically reset upon star-formation?

The spectroscopic signatures of molecules span from the cm through the far-IR, necessitating a breadth of facilities spanning this range. The spatial scales needed to probe these inventories at differing stages from molecular clouds to protoplanetary systems require both single-dish facilities and flexible interferometers (to milliarcsecond scales and below). Spectral lines of complex or low-abundance species can be incredibly weak, requiring exquisite sensitivity (at least 10x ALMA). Many transitions of many molecules must be captured to gain the required picture of chemistry and insight into the underlying astrophysics, necessitating the broadest instantaneous bandwidth possible (i.e. digitizing the entirety of a receiver band in one setting).  Spectral lines range from very narrow (0.1 km/s) to complex in nature and carrying critical kinematic information that can only be extracted with sufficient resolution.  Flexibility to resolve lines at the 0.1 km/s scale is required *without sacrificing bandwidth,* so we must strive to digitize the entirety of receiver bands at better than 0.1 km/s resolution.

**Solar System**
Our Solar System is the most accessible planetary system, a phenomenon we now understand to be ubiquitous in the universe.  What is the structure and composition of the Sun's planets and minor bodies?  How did the Solar System evolve to its current diverse state, and what could this tell us about the formation and evolution of other planetary systems? What is the distribution and origin of volatile species and what processes may influence them?

The far-IR-mm-cm wavelength range is well suited to advance our knowledge of the Solar System, from its small cold distant bodies, to its planets and moons, and even the Sun itself. With increasing wavelength, spectroscopy probes deeper into the atmosphere of a planet. For example, JWST with thermal mid-IR probes the stratosphere and troposphere (from 0.0001 to ~1 bar for Uranus), whereas mm-wave data sample the deep layers (from 0.1 to ~100 bar for Uranus).  Similarly, for moons and Trans Neptunian Objects, longer wavelength data probe increasing depths, as much as meters, below the surface.

Far-IR-mm-cm spectroscopy can also map out the enrichment in heavy elements and variations in isotopic ratios of D/H and of C, N and O, and can monitor for or respond to episodes of outgassing and sublimation.  These data will illuminate the seeding of young Earth and other planets with water, organic molecules and phosphorus, all critical elements for the emergence of life.

The instrumental requirements are very similar to those for Astrochemistry: high sensitivity, high frequency and spatial resolution and large bandwidths. Rapid response remains important for unpredictable transient events.

**Planet Formation**
Planets form and acquire their compositions in disks around young stars. The planets' volatile inventories, including their total gas mass and elemental makeup, all depend on properties of their birth disks. JWST is in the midst of characterizing the volatile inventories of large and mid-sized exoplanets in unprecedented detail, while the upcoming Extremely Large Telescopes (ELT) and the proposed NASA Habitable Worlds

Observatory will characterize potentially habitable terrestrial planets. Thus having an interpretative framework of planet compositions rooted in disk observations for exoplanets is and will remain especially important.

Protoplanetary disks extend from fractions of an au to 100s of au. Current ALMA observations primarily characterize the pebble disk and gas-phase volatiles forming ice giants and comets. JWST and ground-based IR cover the other extreme, the innermost au, where terrestrial planets assemble. These wavelengths also reveal ice features in absorption when the disk geometry allows. Arguably the most important and under-studied disk radii are the intermediate ones, between 1 and 10 au, where most planets (by mass) probably form. To pick out such a region, one must (1) utilize lines with upper state energies that are well coupled to this warm-to-hot gas or (2) directly spatially resolve the gas and dust.

Mapping out the gas mass, structure, and composition at 1-10 au will require combination of spatially and spectrally resolved observations at mm/cm wavelengths (elemental ratios, gas surface density constraints, and organic inventories) and FIR observations (HD, water, and atomic C and O lines). At millimeter wavelengths, present-day ALMA provides sufficient baselines, but lacks sensitivity for gas observations. The sensitivity required to routinely observe the volatile content of the 1-10 au disk region entails increasing the ALMA collecting area by factors of 3 to 10. In terms of the dust, to understand how disks are sculpted by planets in the inner 1-10 au region requires resolved observations enabled by the ngVLA, ideally at high frequencies (30 - 120 GHz), including the CO 1-0 line at 115 GHz to capture inner disk gas flows feeding giant planets.

**Star Formation, Dust and Polarization**
Stars are formed from dense clumps of molecular gas that collapse under gravity while subjected to a number of physical and dynamical processes.  Large quantities of dust in the cloud cores absorb short wavelength light, and re-emit the absorbed energy as thermal emission from dust grains and molecular and atomic transitions.  The FIR and submm regions sample the peak emission from star-forming regions and are key to understanding the detailed processes that lead to the observed distribution of dense cores, disks and stars.  Polarization measurements trace dust grains aligned to magnetic fields and thus provide key indicators of the role of these fields, whilst spectroscopy of key species will unravel the chemical, dynamical and excitation processes that occur.  There are strong indications that the accretion process in protostars is highly time-dependent, and model simulations demonstrate that the accretion behavior of forming stars can be best followed at FIR wavelengths.

**Nearby Galaxies**

Nearby galaxies are the best laboratories for studying the key processes that govern the baryon cycle, because of the high spatial resolution and sensitivity that is possible to attain on them. Some of the important open questions include the formation and properties of interstellar dust and the cold gas phases, the processes that lead to star formation activity in those cold phases and determine locally the fraction of gas mass that is converted into stars and the time scale for such conversion, and the feedback mechanisms that regulate star and black hole formation and consequently galaxy growth on intermediate and large spatial scales. Observations ranging from cm wavelengths (HI) to millimeter, submillimeter, and FIR (including polarimetry) are crucial for addressing these questions.

**Distant Galaxies**

Observations and numerical simulations over the last 10-15 years have demonstrated the important role that cold gas and dust plays in the early formation and evolution of galaxies. Young galaxies at redshift z>1 can have ~50% or more of their baryons in the form of cold interstellar material. For the very early systems (z>5): how did they build their reservoirs of dust? What are the properties of their dust? Are these different from the dust grains found in nearby galaxies? What is the amount of obscured star formation activity? What is their molecular gas content, and their star formation efficiency? How do the first "stable" disks emerge? What is the merger rate? How do the central black holes of these ancient galaxies grow and accrete mass?

For the galaxies at the peak of cosmic star formation activity (or cosmic noon at 1<z<3): How do supermassive black holes and their host galaxies co-evolve? What were the main feedback mechanisms responsible for quenching the star formation activity in massive galaxies? Cosmological starvation, or AGN feedback? How do galaxies acquire the gas to maintain their high levels of star formation activity, and how do they transport this gas to their centers to grow their bulges?

To address those questions, the highest priorities are (1) a 5-10x improvement in sensitivity and greater frequency coverage for ALMA; (2) a FIR probe to observe the rest-frame mid-infrared lines and features at high redshifts; and (3) improved survey capability with a FIR probe or a large aperture (at least 30m) single dish at high altitude observing at mm/submm wavelengths.

**Environments and Ecosystems of Galaxies**

Recent JWST and ALMA studies unveiled the presence of numerous massive systems in the early Universe, such as UV-bright galaxies at $z>\sim 9$, broad-line massive AGNs at $z\sim 8.5$, dusty massive starbursts at z~6.9, and overdensities of massive dusty galaxies at *z*~4.3. These systems are often challenging to interpret with current galaxy formation models. Their physical origins—regulated by star-formation efficiency, feedback, and dark matter assembly—remain a major open question.

To address these questions, the following facilities are important: 1) the ngVLA to achieve high-resolution (< 100pc ~ GMC scales) molecular gas observations for, e.g., low-J CO transitions to investigate the star-formation efficiency in early galaxies, 2) a ~5–10x ALMA and/or 50-m class single-dish mm/submm telescope to recover diffuse extended cold gas such as the [CII]158um transition around early galaxies to directly detect the baryon cycling gas and to probe the related feedback effects, and 3) a wide-field FIR imaging spectrometer to unveil the true abundance of dusty massive systems with the aim of constraining dark matter assembly in the early Universe.

**Time-Domain Astronomy (TDA)**
Time-domain studies of protostars in the FIR are one example of the unique information promised by time-resolved observations at mm, submm and FIR wavelengths. Just as many objects, ISM or IGM phases or physical processes are accessible best or only at mm-to-FIR wavelengths, their time behavior there will also yield unique insights. Time-resolved data have been long known to hold essential astronomical information, but accessible timescales have multiplied dramatically in the visible because of recent technical progress, revealing a wealth of new objects and phenomena from stellar explosions to tidal disruption events. The Astro2020 Decadal Survey made TDA a central recommendation based on that progress. Similarly, variability and transient studies at mm to FIR wavelengths both in the continuum and line emission hold great promise for otherwise inaccessible insights into the cold and dusty universe from planet and star formation to galaxy and black hole evolution. The required capabilities are the facilities described in this report, in a stable network to enable long-term monitoring using short visits.

# Acknowledgments

The Workshop organizers would like to acknowledge the generous support of the Kavli Foundation, and the positive engagement by the IAU Executive Committee during the whole process from early planning to final report generation. The Cahill Center for Astronomy and Astrophysics and IPAC at Caltech hosted and supported the Workshop and Splinter Sessions. They also sponsored Day 2 dinner for participants. The Local Organizing Committee, Frank Aragon, Alice Hang (chair), Debby Miles and Ellen O'Leary, provided invaluable logistical support. This Report summarizes the presentations, discussions and conclusions from the Workshop, and thus reflects input from all participants, written and oral, though space limitations meant that input is not included in its entirety. The Report was authored and edited by the Science Organizing Committee and benefited greatly from copy editing by Alice Hang.

# Appendix A.  Technical considerations for submm telescopes and instruments

Telescopes operating at submm/mm wavelengths can operate either as stand-alone facilities or linked together to form an interferometer. The spatial resolution in these two cases is very different, with the image resolution depending on the diameter of the telescope aperture in the first case, and on the separation of the telescope elements in the second. The instruments at the two kinds of facilities are also different, with the greatest performance in a single dish being obtained with detector arrays that sample large areas of the telescope focal plane, whilst interferometers currently have single (or small numbers of) receiver pixels but operate at both high spatial and spectral resolution. In general, interferometers are well suited to studying objects with sizes or structures similar to the instrument resolution, whereas single dish facilities are more effective when surveying or mapping larger areas or low-surface brightness regions. Although the capability of interferometers for measuring larger spatial scales can be enhanced by inclusion of smaller telescopes, e.g., the 12 7-meter antennas in the ALMA Compact Array, the resulting mapping speed falls far short of that now achievable using single-dish telescopes with large focal-plane arrays.

A relevant metric for mapping speed is the product of the throughput or étendue and the detection bandwidth, $A\Omega\Delta\nu$, where $A$ is the telescope collecting area and $\Omega$ is the instantaneous field of view sampled by the detectors. To within factors of order unity, the normalized throughput $A\Omega/\lambda^2$ for an interferometric array is just the number of dishes $N_d$ (66 for ALMA currently) while for a single-dish telescope it is the number of pixels $N_p$ in the focal plane. For single-dish telescopes, innovative technologies have led to dramatic increases in detector sensitivity, manufacturability, and array formats. Indeed, recent developments in kinetic inductance detector (KID) arrays enable mm/submm instruments with $N_p = 10^5$-$10^6$ pixels, with per-pixel detection bandwidths approaching $\Delta\nu = 100$ GHz, limited by the available atmospheric windows. The gains produced with detector arrays that can potentially fill the focal plane are enormous, with huge increases in mapping speed and area. Meanwhile advances in electronics and computational capability allow increased bandwidths for interferometers; for example, the ALMA [Wideband Sensitivity Upgrade](.) (WSU) aims to double the detection bandwidth to $\Delta\nu = 16$ GHz per polarization. For continuum point-source sensitivity, the relevant metric is the product of collecting area and bandwidth, $A\Delta\nu$. Thus, ALMA's current total collecting area with $\Delta\nu = 16$ GHz bandwidth is equivalent to a 53 m diameter single dish with $\Delta\nu = 50$ GHz bandwidth. A 50 m AtLAST would therefore provide continuum sensitivity comparable to ALMA/WSU but with >$10^3$ times faster mapping speed.

As illustrated in Figure A-1, existing mm-wave single-dish telescopes are equipped with array cameras and receivers. Yet, existing instruments do not come close to pushing the limits of what is now technologically feasible. For example, the 6 m CCAT-prime observatory now under construction aims to field a multi-band wide-field camera comprising nearly $10^5$ KID pixels, and a several megapixel multi-band camera is envisioned as a second-generation instrument for AtLAST.

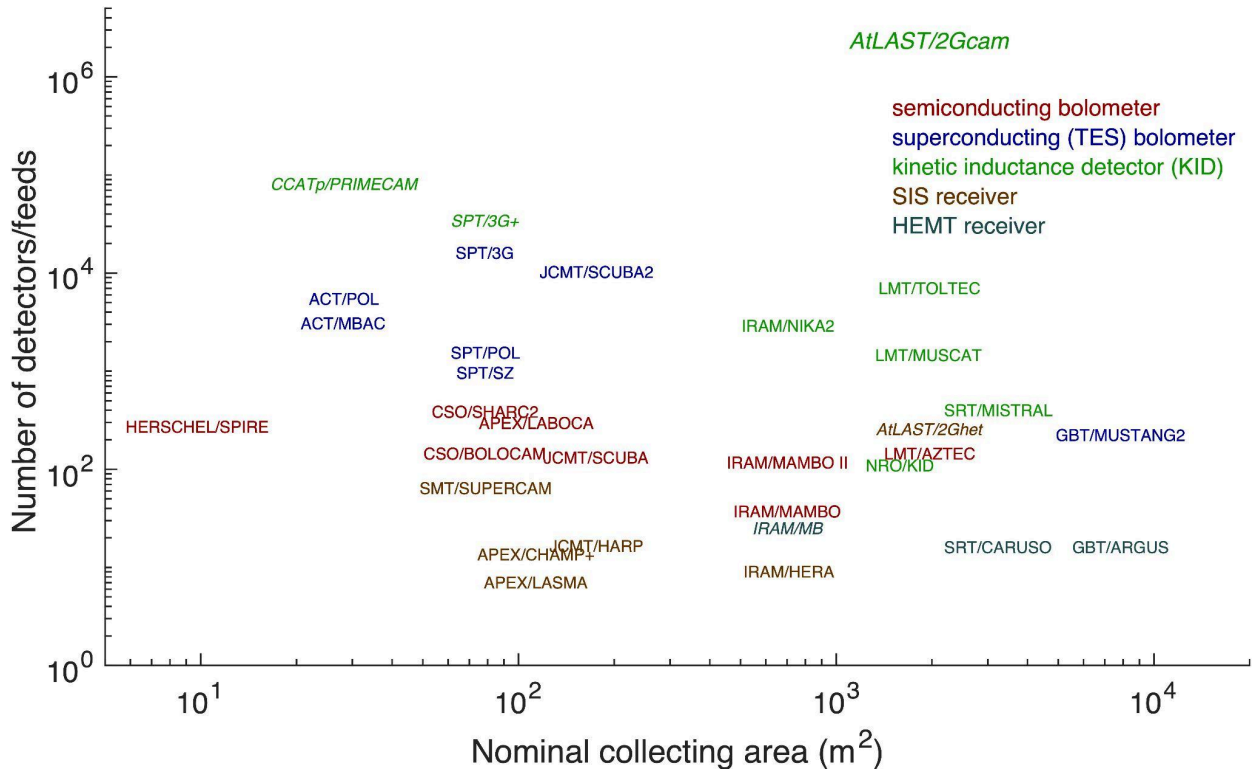

Figure A-1: Array instruments for mm-wave single-dish telescopes; planned capabilities are shown in italic. The technological progression of continuum cameras from semiconductor bolometers to TES to KID is clearly apparent. Array sizes for coherent spectroscopic receivers (SIS, HEMT) are significantly smaller than for the continuum cameras. Note that this figure omits important factors such as observing frequency/wavelength, aperture efficiency, site quality, polarization capability, etc.

# Appendix B. List of Workshop Attendees

Itziar Aretxaga — INAOE
Edwin Anthony Bergin — University of Michigan
Geoffrey A Blake — California Institute of Technology
Alberto D. Bolatto — University of Maryland, College Park
Geoffrey Bower — ASIAA
Charles M Bradford — JPL / Caltech
John Carpenter — Joint ALMA Observatory
Scott Chapman — Dalhousie University
James Okwe Chibueze — University of South Africa
Claudia Cicone — University of Oslo
Lauren Ilsedore Cleeves — University of Virginia
Asantha Cooray — UC Irvine
Pierre Cox — Institut d'Astrophysique de Paris
Daniel Dale — University of Wyoming
Katherine De Kleer — Caltech
Akira Endo — TU Delft
Stefano Facchini — University of Milan
Seiji Fujimoto — UT Austin
Matt J Griffin — Cardiff University
Daniel Harsono — National Tsing Hua University
Bert Hawkins — NRAO
George Helou — Caltech
Thomas Kai Henning — MPI for Astronomy
Brandon Hensley — JPL/Caltech
Rodrigo Herrera Camus — Universidad de Concepción
Lynne Hillenbrand — Caltech
Luis C Ho — Kavli Institute for Astronomy and Astrophysics, Peking University
Elizabeth Humphreys — ESO
Satoru Iguchi — National Astronomical Observatory of Japan
Hanae Inami — Hiroshima University
Doug Johnstone — NRC-Herzberg

Robert Kennicutt — U Arizona/TAMU

Kotaro Kohno — University of Tokyo

Guilaine Lagache — LAM

David T Leisawitz — NASA / GSFC

Enrique Lopez Rodriguez — KIPAC at Stanford University

Meredith Ann Mac Gregor — Johns Hopkins University

Gregory Mack — The Kavli Foundation

Brett Mc Guire — MIT

Margaret Meixner — Jet Propulsion Laboratory, California Institute of Technology

Kentaro Motohara — National Astronomical Observatory of Japan

Eric Joseph Murphy — NRAO

Karin Oberg — Harvard University

Klaus Pontoppidan — Jet Propulsion Laboratory

Alexandra Pope — University of Massachusetts Amherst

Laura Pérez — Universidad de Chile

Patrick F Roche — Oxford University

Monica Rubio — Departamento de Astronomia, Universidad de Chile

Karin Sandstrom — UC San Diego

Anneila Sargent — Caltech

Marc Sauvage — CEA/DRF/Irfu/DAp

Eva Schinnerer — MPIA

Bernhard Schulz — DSI, University Stuttgart

Karl Friedrich Schuster — IRAM

Nick Scoville — Caltech

John David Smith — Toledo

Justin Spilker — Texas A&M University

Tarun Souradeep — Raman Research Institute

Gordon Stacey — Cornell University

Amiel Sternberg — Tel Aviv University

Eric Switzer — NASA Goddard

Linda Tacconi — Max-Planck-Institute for Extraterrestrial Physics

Yoichi Tamura — Nagoya University

Ewine Van Dishoeck — Leiden University

Joaquin Vieira — U. Illinois / NCSA

Christopher K Walker — University of Arizona
Fabian Walter — MPIA Heidelberg
Michael Werner — JPL
Jonas Zmuidzinas — Caltech

# Appendix C. Workshop Program

## Kavli-IAU Workshop on Global Coordination
## Probing the Universe From Far-infrared to Millimeter Wavelengths: Future Facilities and their Synergies

*Hameetman Auditorium, Cahill Center for Astronomy and Astrophysics, Caltech*

*Talk lengths: 40=30+10 and 30=25+5*
*Panels start with 5 minutes/panelist followed by Q&A with audience*

### Tuesday, March 26, 2024

| Time | Title | Speaker |
|---|---|---|
| 8:00 AM | Morning Coffee and Tea | |
| 8:30 AM | Welcome and Plan of Workshop | SOC |
| 9:00 AM | Large Facilities: Landscape to 2040 | Ewine van Dishoeck |
| 9:45 AM | ALMA: Today Through 2040 | Liz Humphreys |
| 10:25 AM | Break | |
| 10:45 AM | Ground-based submm: Panel Discussion (Moderator: Ilse Cleeves) | Geoffrey Bower, John Carpenter, Bert Hawkins, Satoru Iguchi, Laura Perez, Karl Schuster |
| 12:00 PM | Lunch (Box lunch provided) | |
| 1:00 PM | Far-IR Space Missions | Matt Griffin |
| 1:40 PM | Far-IR Missions: Panel Discussion (Moderator: Kentaro Motohara) | Hanae Inami, Guilaine Lagache, Meredith MacGregor, Alex Pope, Marc Sauvage, Justin Spilker, Eric Switzer |
| 2:55 PM | Break | |
| 3:15 PM | ngVLA | Eric Murphy |
| 3:45 PM | CMB-S4 | Joaquin Viera |
| 4:15 PM | Break | |
| 4:35 PM | Ground-based Projects: Panel Discussion (Moderator: Alberto Bolatto) | Itziar Aretxaga, Scott Chapman, Claudia Cicone, Akira Endo, Yoichi Tamura |
| 5:50 PM | Adjourn | |
| 6:00 PM | Mingle and Dinner in Dabney Lounge | |

### Wednesday, March 27, 2024: Science Opportunities

| Time | Title | Speaker |
|---|---|---|
| 8:00 AM | Morning Coffee and Tea | |
| 8:30 AM | Announcements | SOC/LOC |
| 8:40 AM | Astrochemistry | Brett McGuire |
| 9:10 AM | Star Formation | Thomas Henning |
| 9:40 AM | Disks | Karin Oberg |
| 10:10 AM | Break | |
| 10:30 AM | Planet Formation + Exoplanets | Stefano Facchini |
| 11:00 AM | Solar System | Katherine de Kleer & Stefanie Milam |
| 11:30 AM | Time Domain | Doug Johnstone |

# Kavli-IAU Workshop on Global Coordination
# Probing the Universe From Far-infrared to Millimeter Wavelengths:
# Future Facilities and Their Synergies

*Hameetman Auditorium, Cahill Center for Astronomy and Astrophysics, Caltech*

*Talk lengths: 40=30+10 and 30=25+5*
*Panels start with 5 minutes/panelist followed by Q&A with audience*

## Wednesday, March 26, 2024: Science Opportunities (continued)

| Time | Title | Speaker |
|---|---|---|
| 12:00 PM | Lunch (Box lunch provided) | |
| 1:00 PM | ISM At All Scales: Panel Discussion (Moderator: Pat Roche) | Ted Bergin, Brandon Hensley, James Chibueze, Monica Rubio, Daniel Harsono, Amiel Sternberg |
| 2:15 PM | Break | |
| 2:35 PM | Galaxies Not So Far | Karin Sandstrom |
| 3:05 PM | Galaxies Very Far | Rodrigo Herrera-Camus |
| 3:35 PM | Galaxies: Environments and Ecosystems | Seiji Fujimoto |
| 4:05 PM | Break | |
| 4:25 PM | Galaxies and Cosmology: Panel Discussion (Moderator: Daniel Dale) | Kotaro Kohno, Enrique Lopez-Rodriguez, Eva Schinnerer, JD Smith, Gordon Stacey |
| 5:40 PM | Adjourn | |
| 6:00 PM | Mingle and Dinner at the Athenaeum | |

## Thursday, March 28, 2024: The Path Forward

| Time | Title | Speaker |
|---|---|---|
| 8:00 AM | Morning Coffee and Tea | |
| 8:30 AM | Organization and Logistics | SOC |
| 8:45 AM | Break-out Sessions<br>• High-Res: ALMA, ngVLA<br>• Deep: Far-IR Space<br>• Broad: CMS-S4, other ground | SOC & TBD |
| 10:15 AM | Break | |
| 10:45 AM | Break-out Session Reports | SOC |
| 11:30 AM | Group Discussion (Moderator: Ewine van Dishoeck) | All |
| 12:30 PM | SOC Plan Forward | SOC |
| 12:45 PM | Adjourn Main Workshop and Lunch (Box lunch provided) | |
| 1:45 PM | SOC Work Session to assemble key points/conclusions/recommendations, draft report | |
| 4:45 PM | Adjourn SOC Work Session | |

# Appendix D.  Acronym List

| Acronym | Definition |
| --- | --- |
| ACT | Atacama Cosmology Telescope |
| AGN | Active Galactic Nuclei |
| ALMA | Atacama Large Millimeter/ submillimeter Array |
| APEX | Astrophysics Probe Explorer |
| APEX | Atacama Pathfinder Experiment |
| AS | Academia Sinica |
| AtLAST | Atacama Large Aperture Submillimeter Telescope |
| CCAT | Cerro Chajnantor Atacama Telescope |
| CBI | Cosmic Background Imager |
| CDR | Conceptual Design Review |
| CMB | Cosmic microwave background |
| CSO | Caltech Submillimeter Observatory |
| DASI | Degree Angular Scale Interferometer |
| DESHIMA | Deep Spectroscopic High-redshift Mapper |
| ELT | Extremely Large Telescopes |
| ESO | European Southern Observatory |
| FIR | Far-infrared |
| FIRSST | Far-Infrared Spectroscopic Survey Telescope |
| FTS | Fourier transform spectrograph |
| IAU | International Astronomical Union |
| GA | General Assembly |
| GBT | Green Bank Telescope |
| GMC | Giant Molecular Cloud |
| HAWC+ | High-resolution Airborne Wide-Band Camera |
| HEMT | High Eelctron Mobility Transistor |
| IR | Infrared |
| IRAM | Institut de Radioastronomie Millimetrique |
| ISM | Interstellar medium |
| ISO | Infrared Space Observatory |
| JCMT | James Clerk Maxwell Telescope |
| JWST | James Webb Space Telescope |
| JVLA | Jansky Very Large Array |
| KASI | Korea Astronomy and Space Science Institute |
| KIDs | Kinetic inductance detectors |
| LMT | Large Millimeter Telescope |
| LST | Large Submillimeter Telescope |
| MREFC | Major Research Equipment and Facilities Construction |
| MSIP | Mid-Scale Innovation Program |
| ngVLA | Next Generation Very Large Array |
| NASA | National Aeronautics and Space Administration |
| NINS | National Institutes of Natural Sciences (Japan) |
| NIRSpec | Near Infrared Spectrograph |
| NOEMA | Northern Extended Millimeter Array |
| NRAO | National Radio Astronomy Observatory |
| NRC | National Research Council (Canada) |
| NRO | Nobeyama Radio Observatory |
| NSF | National Science Foundation |

| OST | Origins Space Telescope |
| --- | --- |
| PDR | Preliminary Design Review |
| PRIMA | PRobe Infrared Mission for Astrophysics |
| SALTUS | Single Aperture Large Telescope for Universe Studies |
| SED | Spectral Energy Distribution |
| SIS | Superconductor-insulator-superconductor |
| SKA | Square Kilometre Array |
| SMA | Submillimeter Array |
| SMT | Submillimeter Telescope |
| SOFIA | Stratospheric Observatory for Infrared Astronomy |
| SPICA | Space Infrared Telescope for Cosmology and Astrophysics |
| SPIRE | Spectral and Photometric Imaging Receiver (Herschel) |
| SPT | South Pole Telescope |
| SRT | Sardinia Radio Telescope |
| TDA | Time-Domain Astronomy |
| TES | Transition Edge Sensor |
| UV | Ultraviolet |
| VIPA | Virtual Imaged Phased Array |
| VLBI | Very Long Baseline Interferometry |
| VLBA | Very Large Baseline Array |
| WG | Working Group |
| WSU | Wideband Sensitivity Upgrade (ALMA) |